\documentclass[letterpaper, 10 pt, conference]{ieeeconf}
\IEEEoverridecommandlockouts 

\usepackage{cite}
\usepackage{amsmath, amssymb, amsfonts}
\usepackage{graphicx}
\usepackage{textcomp}
\usepackage{xcolor}
\usepackage{algorithm}
\usepackage{algpseudocode}
\usepackage{tikz}
\usetikzlibrary{shapes, positioning}
\usepackage{longtable}
\usepackage{multirow}
\usepackage{etoolbox}
\usepackage{hyperref} 
\usepackage{subfig}
\usepackage{subcaption}
\usepackage{mathtools}

\usepackage{booktabs}

\usepackage{enumerate}

\usepackage{bm}

\newtheorem{proposition}{Proposition}
\newtheorem{lemma}{Lemma}

\newtheorem{remark}{Remark}
\newtheorem{definition}{Definition}
\newtheorem{corollary}{Corollary}

\AtEndEnvironment{problem}{\hfill$\Box$}
\AtEndEnvironment{proposition}{\hfill$\Box$}
\AtEndEnvironment{lemma}{\hfill$\Box$}
\AtEndEnvironment{theorem}{\hfill$\Box$}
\AtEndEnvironment{assumption}{\hfill$\Box$}
\AtEndEnvironment{condition}{\hfill$\Box$}
\AtEndEnvironment{remark}{\hfill$\Box$}
\AtEndEnvironment{corollary}{\hfill$\Box$}
\AtEndEnvironment{definition}{\hfill$\Box$}

\newcommand{\R}{\mathbb{R}}
\newcommand{\norm}[1]{\left|#1\right|}
\DeclareMathOperator{\Span}{\text{span}}

\providecommand{\SO}{\mathbf{SO}}

\begin{document}

\title{\LARGE \bf Scalar-Measurement Attitude Estimation on $\SO(3)$ with Bias Compensation}

\author{A.~Melis, T.~Bouazza, H.~Alnahhal, S.~Benahmed,\\
S.~Berkane,~\IEEEmembership{Senior Member,~IEEE}, 
and T.~Hamel,~\IEEEmembership{Fellow,~IEEE}
\thanks{Alessandro Melis, Tarek Bouazza, and Tarek Hamel are with I3S, CNRS, Université Côte d'Azur, Sophia Antipolis, France. Tarek Hamel is also with the Institut Universitaire de France
        (\{melis,bouazza,thamel\}@i3s.unice.fr).}
\thanks{Hassan Alnahhal is an independent researcher based in Cairo, Egypt (h.alnahhal1989@gmail.com).}
\thanks{Sifeddine Benahmed is with the Department of Technology \& Innovation, Capgemini Engineering, Toulouse, 31300, France (sif-eddine.benahmed@capgemini.com).}
\thanks{Soulaimane Berkane is with the Department of Computer Science and Engineering, Université du Québec en Outaouais (UQO), QC J8X 3X7, Canada (soulaimane.berkane@uqo.ca).}
\thanks{*This research is supported in part by the ASTRID ANR project ASCAR, the ``Grands Fonds Marins'' Project Deep-C, the Natural Sciences and Engineering Research Council of Canada (NSERC) under the Discovery Grant RGPIN-2020-04759, and the Fonds de recherche du Québec (FRQ).}
}
\maketitle

\begin{abstract}
Attitude estimation methods typically rely on full vector measurements from inertial sensors such as accelerometers and magnetometers. This paper shows that reliable estimation can also be achieved using only scalar measurements, which naturally arise either as components of vector readings or as independent constraints from other sensing modalities. We propose nonlinear deterministic observers on $\SO(3)$ that incorporate gyroscope bias compensation and guarantee uniform local exponential stability under suitable observability conditions. A key feature of the framework is its robustness to partial sensing: accurate estimation is maintained even when only a subset of vector components is available. Experimental validation on the BROAD dataset confirms consistent performance across progressively reduced measurement configurations, with estimation errors remaining small even under severe information loss. To the best of our knowledge, this is the first work to establish fundamental observability results showing that two scalar measurements under suitable excitation suffice for attitude estimation, and that three are enough in the static case. These results position scalar-measurement-based observers as a practical and reliable alternative to conventional vector-based approaches.

\end{abstract}


\section{Introduction}
Accurate attitude (orientation) estimation is a fundamental requirement in diverse applications such as spacecraft stabilization, aerial and ground vehicle control, and autonomous navigation \cite{crassidis2007survey}. To achieve this, reliable state observers (filters) have been extensively studied, aiming to fuse information from multiple sensors—typically providing vector inertial measurements, expressed in either the inertial or body frame, together with body-frame angular velocity measurements \cite{Mahony_Hamel_Pflimlin,batista2012sensor,zlotnik2016nonlinear,tayebi2006attitude,izadi2014rigid}. The most widely used sensors for attitude estimation are Inertial Measurement Units (IMUs), which typically integrate tri-axial accelerometers and gyroscopes, and may also include a magnetometer. Despite their low cost and compactness, IMUs suffer from limited accuracy and high sensitivity to noise. In particular, gyroscopes—crucial for observer design due to their high bandwidth—are affected by drift and bias \cite{lefferts1982kalman}.

Early solutions to Wahba’s problem \cite{wahba1965least} focused on deterministic reconstruction methods, such as Davenport’s $q$-method, Shuster’s QUEST algorithm, and Markley’s SVD-based approach \cite{shuster1981three,markley1988attitude}. These techniques provide closed-form or iterative solutions to the attitude estimation problem from vector observations, but they do not explicitly handle sensor noise or dynamic models. To address these aspects, Kalman-type filters were later introduced for attitude and bias estimation \cite{lefferts1982kalman}. However, these filters are computationally intensive and rely on linear approximations, requiring careful tuning and implementation~\cite{crassidis2007survey}. To overcome such limitations, invariant Kalman filters have emerged as a robust alternative, offering local asymptotic stability and improved performance in nonlinear settings~\cite{barrau2018invariant,bonnabel2009invariant}. In parallel, nonlinear deterministic observers have been developed, offering stronger theoretical stability guarantees and better handling of the nonlinear dynamics inherent to Inertial Navigation Systems (INS)~\cite{thienel2001coupled,hua2014implementation,zlotnik2016nonlinear,Mahony_Hamel_Pflimlin,grip2011attitude,berkane2017design}. While these approaches have shown promising results, they typically assume the availability of complete 3D vector measurements--an assumption that may not hold in practice~\cite{alnahhal2025scalar}.

This paper revisits the scalar-measurement framework for rigid-body orientation estimation first introduced in~\cite{alnahhal2025scalar}. Such measurements are often expressed as the cosine of the angle between two vectors, but they may also arise from alternative constraints, including tilt relations obtained from barometer and range sensors or landmark-based altitude differences. Unlike full vector observations, scalar measurements provide only partial constraints on the attitude, yet they enable a more flexible and robust estimation process, particularly in scenarios where directional data are sparse, noisy, or partially unavailable. The formulation in~\cite{alnahhal2025scalar}, however, relied on embedding $\SO(3)$ into $\mathbb{R}^9$, which imposed unnecessarily strong uniform observability conditions and did not address bias. To overcome these limitations, we adopt a direct design on $\SO(3)$ that explicitly incorporates gyroscope bias. Building on the Riccati-based observer framework of~\cite{hamel2018riccati}, we propose a deterministic filter inspired by the multiplicative Extended Kalman Filter (MEKF), where the continuous Riccati equation (CRE) governs both observer dynamics and Lyapunov-based stability analysis~\cite{hamel2017position}.

The main contributions are fourfold: (i) a direct $\SO(3)$-based observer that avoids high-dimensional embeddings, (ii) explicit bias handling within a deterministic Riccati framework, (iii) rigorous persistence-of-excitation conditions linking observability of the linearized dynamics to local exponential stability, and (iv) new fundamental insights into scalar-based estimation. In particular, while it is well known that at least two non-collinear vector measurements are required to reconstruct attitude~\cite{Mahony_Hamel_Pflimlin,batista2012sensor}, we show that two scalar measurements under suitable excitation suffice for attitude observability, and that three scalar measurements are enough in the static case, matching the intrinsic three-dimensional nature of $\SO(3)$. To the best of our knowledge, these results have not been reported before. Finally, the proposed approach is validated experimentally under partial and noisy measurements.

The remainder of this paper is organized as follows. Section~II introduces the notation, system equations, and measurement models, along with essential concepts on uniform observability and the CRE-based observer. Section~III presents the observer design for both unbiased and biased angular velocity measurements, together with conditions ensuring local exponential stability. Section~IV reports experimental results, and Section~V concludes the paper.

\section{Preliminary Material}
\subsection{Notation}
\begin{itemize}
    \item $\mathcal{I} = \{G_{\mathcal{I}}, e_1, e_2, e_3\}$ denotes a right-handed inertial reference frame with fixed origin $G_\mathcal{I}$ and standard basis vectors of $\mathbb{R}^3$. The body-fixed frame $\mathcal{B} = \{G_\mathcal{B}, e_{B1}, e_{B2}, e_{B3}\}$ is attached to the vehicle, with its origin $G_\mathcal{B}$ at the center of mass.
    
    \item The Euclidean norm of a vector $x \in \mathbb{R}^n$ is denoted by $|x|$.
    The unit 2-sphere is represented as $\mathbb{S}^2 := \{v \in \mathbb{R}^3 \mid |v| = 1\}$.
    The set $\mathfrak{B}^n_r := \{ x \in \R^n \mid |x| \leq r \}$ denotes the closed ball in $\R^n$ of radius $r$.
    
    \item For any vector $\Omega \in \mathbb{R}^3$, $\Omega^\times$ is the skew-symmetric matrix associated with the cross product, satisfying $\Omega^\times y = \Omega \times y$ for all $y\in \R^3$.

    \item The special orthogonal group, denoted as $\SO(3)$, is the Lie group of 3D rotations given by
    $\SO(3) := \{R \in \mathbb{R}^{3\times3} \mid R^{\top} R = RR^\top = I_3, \det(R) = 1\}$.
    Its associated Lie algebra is defined as
    $\mathfrak{so}(3) := \{\Omega^\times \mid \Omega \in \mathbb{R}^3\}$.

    \item Let $f$ be a vector-valued function of two variables $x$ and $y$, and time $t$. We write  
    $f = O(|x|^{k_1} |y|^{k_2})$, with $k_1 \geq 0, k_2 \geq 0,
    $
    if for all $t$: $|f(x, y, t)|/(|x|^{k_1} |y|^{k_2}) \leq \gamma < \infty$ in the vicinity of $(x = 0, y = 0)$.
\end{itemize}
For clarity and conciseness, the argument of time-dependent signals is provided when necessary and omitted otherwise.
\subsection{Observability Principles and Requirements}
Consider a generic linear time-varying (LTV) system:
\begin{subequations} \label{eq:LTV_system_dynamics}
    \begin{align}
        \dot{\bm{x}} &= A(t) \bm{x} + B(t) \bm{u}, &
       \bm{y} &=C(t) \bm{x},
    \end{align}
\end{subequations}
where \( \bm{x} \in \mathbb{R}^n \) denotes the state, \( \bm{u} \in \mathbb{R}^s \) the input, and \( \bm{y} \in \mathbb{R}^m \) the output. The following definition of observability related to this system is derived from the works  \cite{aeyels1998asymptotic,chen1984linear}.

\begin{definition}[Uniform Observability] \label{def:uniform_obs}
The system \eqref{eq:LTV_system_dynamics} is said to be \textit{uniformly observable} if there exist constants $\delta, \mu > 0$ such that, for all $t \geq 0$:  
\begin{equation}  
W(t, t + \delta) \coloneqq \frac{1}{\delta} \int_{t}^{t+\delta} \Phi^\top(s, t)C^\top(s)C(s)\Phi(s, t) \, ds \succeq \mu I_n, 
\label{eq:main Uniform Observability condition}
\end{equation}  
where $W(t, t + \delta)$
is the observability Gramian associated with the pair $(A(t), C(t))$, and  
$\Phi(s, t)$ denotes the state transition matrix associated with $A(t)$, defined by: $\frac{d}{dt} \Phi(s, t) = A(t) \Phi(s, t)$, $\Phi(t, t) = I_n$.
If condition \eqref{eq:main Uniform Observability condition} holds, the pair $(A(t), C(t))$ is said to be \textit{uniformly observable}.
\end{definition}

\subsection{Riccati Observers for a Class of Nonlinear Systems}
Consider a class of nonlinear systems with state 
$\bm{x} := (x_1, \, x_2) \in \mathfrak{B}^{n_1}_r\times \R^{n_2}$, $n = n_1 + n_2$, input $\bm{u}\in \R^s$, and output $\bm{y}\in\R^m$ satisfying the dynamics
\begin{subequations}\label{eq:riccati_dynamics}
    \begin{align}
    \dot{\bm{x}} &= A(x_1,t)\bm{x} + \bm{u} + O(|x_1||\bm{u}|) + O(|x_1||x_2|), \\
    \bm{y} &= C(x_1,t) \bm{x} + O(|x_1|^2),
    \end{align}
\end{subequations}
where the matrix-valued functions $A(x_1,t) \in \R^{n \times n}$ and $C(x_1,t) \in \R^{m \times n}$ are continuous matrix-valued functions uniformly bounded w.r.t. $t$. 
The following result is adapted from Theorem 3.1 and Corollary 3.2 in \cite{hamel2018riccati}.
\begin{proposition}\label{proposition: uniform observability to stability}
    Consider the system dynamics \eqref{eq:riccati_dynamics} and choose the input signal:
    \begin{equation}\label{eq:riccati_input}
        \bm{u} = -P(t) C^\top(x_1,t) Q(t) \bm{y},
    \end{equation}
    with $P(0)$ a positive definite (p.d.) matrix solution to the Continuous Riccati Equation (CRE)
    \begin{equation}
    \begin{aligned}
        \dot{P} = &A P + P A^\top - PC^\top Q(t) C P + V(t),
    \end{aligned}
        \label{eq:CRE}
    \end{equation}
    with $Q(t)$ and $V(t)$ bounded continuous symmetric positive definite matrix-valued functions.
    If the pair $(A^\star(t),C^\star(t)):=(A(0,t),C(0,t))$ is uniformly observable, then the origin of \eqref{eq:riccati_dynamics} is locally exponentially stable.
\end{proposition}
The result follows from standard Riccati-based Lyapunov arguments. 
Since $A(x_1,t)$ and $C(x_1,t)$ depend only on $x_1$, compactness of $\mathfrak{B}^{n_1}_r$ ensures their uniform boundedness along trajectories. 
This, combined with uniform observability of the pair $(A^\star(t),C^\star(t))$, guarantees boundedness and well-conditioning of $P(t)$, from which local exponential stability follows. 
See Theorem 3.1 in \cite{hamel2018riccati} for details (cf. \cite{Melis2026}).

\subsection{System Equations and Measurements}
Let $R \in \SO(3): \mathcal{B} \rightarrow \mathcal{I}$ denote the orientation of a moving body, mapping the body-fixed frame $\mathcal{B}$ to the inertial frame $\mathcal{I}$. The attitude evolves according to  $\dot{R} = R \Omega^\times$, 
where $\Omega \in \mathbb{R}^3$ is the angular velocity, typically provided by a tri-axial gyroscope. In practice, the gyro measurement takes the form  $\Omega_y = \Omega + d + \mu_\Omega$
where the true body-frame angular velocity is corrupted by an additive noise term $\mu_\Omega$ and a slowly time-varying bias $d$. Since we focus on the design of a deterministic observer, we idealize the measurement by assuming noise-free gyros ($\mu_\Omega \equiv 0$), and adopt the following model to describe the dynamics of the system of interest:
\begin{align}
        \dot{R} &= R (\Omega_y - d)^\times, &
        \dot d &= 0.
\end{align}\label{eq:orientation_dynamic}
As for the output measurements, the moving object is assumed to be equipped with a suite of sensors located at the origin of $\mathcal{B}$, providing the outputs $y_{i} \in \mathbb{R}^{n_i}$, $i=1,\cdots,p$, defined as
\begin{equation}\label{eq:general_output}
    y_i := \Lambda_i^\top R^\top b_i, \qquad \Lambda_i:=\begin{bmatrix}
        a_1^{(i)} &\cdots& a_{n_i}^{(i)}
    \end{bmatrix},
\end{equation}
which compose the full output measurement vector 
\begin{equation*}
    y:=[y_1^\top \ \cdots\ y_p^\top]^\top\in \R^m,
\end{equation*} with $m = \sum_{i=1}^pn_i$. Each $y_i$ represents the set of scalar outputs associated with the possibly time-varying known inertial direction $b_i \in \mathbb{R}^3$, collected by measuring $b_i$ along the body-frame directions ${a}_j^{(i)} \in \mathbb{R}^3$, $j = 1,\cdots, n_i$, $n_i \in \mathbb{N}/\{0\}$. 
In the case of a classical IMU, let $R_{\mathrm{IMU}}\in\SO(3)$ denote the constant alignment between the IMU sensor axes and the body frame $\mathcal{B}$.  
Assuming inertial acceleration is negligible compared to gravity, the accelerometer provides three scalar measurements collected in the output
$y_1 = \Lambda_1^\top R^\top b_1$ with $\Lambda_1\in\R^{3\times 3}$ composed by $a_j^{(1)} = R_{\mathrm{IMU}}e_j$, for $j=1,2,3$, and $b_1 = e_3$, where $e_3$ is the gravity direction in the inertial frame.  
Similarly, the magnetometer channels satisfy the same model with $a_j^{(2)} = R_{\mathrm{IMU}}e_{j}$, for $j=1,2,3$, and $b_2 = m_0$, where $m_0$ denotes the magnetic field direction in the inertial frame.  
Further examples of scalar attitude measurements derived from other sensors can be found in~\cite{alnahhal2025scalar}.

\section{Observer Design} \label{sec:observer_design}
\begin{figure}[t]
\centerline{\includegraphics[width=.9\linewidth]{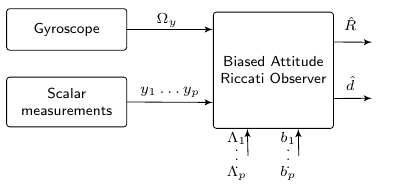}}
\vspace{-.3cm}
\caption{Illustration of the proposed estimation approach.}
\label{fig:estimation_approach}
\end{figure}

By exploiting the above modeling \eqref{eq:orientation_dynamic}-\eqref{eq:general_output}, the proposed observer equation is posed directly as a kinematic system for an attitude estimate on $\SO(3)$ along with an estimate of the bias in $\mathbb{R}^3$.  
The observer kinematics consist of a prediction term based on the measurement $\Omega_y$, and an innovation term $\Delta = (\Delta_R, \Delta_d) \in \R^6$ derived from the available measurements through the Riccati observer framework described in \cite{hamel2018riccati}. 
The proposed attitude estimation approach is illustrated in Fig.~\ref{fig:estimation_approach}.
The general form proposed for the observer is:
\begin{subequations}\label{eq:biased_obs_dyn}
\begin{align}
        \dot{\hat R} &= \hat R(\Omega_y - \hat d)^\times + \Delta_R^\times \hat{R},\quad\hat{R}(0)=\hat{R}_0,  \\
        \dot{\hat d} &= -\Delta_d,\quad \hat{d}(0)=\hat{d}_0.
\end{align}
\end{subequations}
Let $\tilde d = d - \hat d \in \mathbb{R}^3$ denote the bias error, and define the attitude error on $\SO(3)$ by $\tilde{R} = \hat{R}R^\top$. Combining the true and estimated dynamics yields the nonlinear error dynamics:
\begin{align}\label{eq:attitude_error_dynamic}
        \dot{\tilde R} &= \left(\hat{R}\tilde d + \Delta_R\right)^\times \tilde R, &
        \dot{\tilde d} &= \Delta_d.
\end{align}

The output error is defined as
\begin{equation}
\tilde{y}_i := \hat{y}_i - y_i   
=  \Lambda_i^\top \hat{R}^\top b_i - \Lambda_i^\top R^\top b_i
= \Lambda_i^\top \hat{R}^\top (I_3 - \tilde{R}) b_i.
\label{eq:output_error}
\end{equation}

To derive a first-order approximation of the error dynamics \eqref{eq:attitude_error_dynamic}, we exploit a local compact parametrization of $\SO(3)$ via the unit quaternion $\tilde{Q} := (\tilde{q}_0, \tilde{q}) \in \mathbb{S}^3$ corresponding to the rotation error $\tilde{R}$. Rodrigues’ formula writes
\begin{equation}
   \tilde{R} = I_3 + 2\tilde{q}^\times (\tilde{q}_0 I_3 + \tilde{q}^\times),
\end{equation}
from which a first-order approximation of $\tilde{R}$ around $I_3$ is obtained as
\begin{equation}
\tilde{R} = I_3 + \tilde{\lambda}^\times + O(|\tilde{\lambda}|^2),
\label{eq:approx_R_error_with_lambda}
\end{equation}
with $\tilde{\lambda} := 2 \operatorname{sign}(\tilde{q}_0) \tilde{q} \in \mathfrak{B}_2^3$.  
It follows that the observer error dynamics can be expressed as
\begin{subequations}\label{eq:observer_dynamics}
\begin{align}
        \dot{\tilde \lambda} &= \hat R \tilde d + \Delta_R 
        + O(|\tilde{\lambda}||\Delta_R|\big) + O\big(|\tilde{\lambda}||\tilde{d}|), \\
        \dot{\tilde d} &= \Delta_d,
\end{align}
\end{subequations}
while the output error becomes
\begin{equation}
\tilde{y}_i = \Lambda_i^\top \hat R b_i^\times \tilde{\lambda} 
+ O(|\tilde{\lambda}|^2). 
\label{eq:observer_output}
\end{equation}
By setting 
\( \bm{x} := (\tilde{\lambda}, \, \tilde{d})\in \mathfrak{B}_2^3\times\mathbb{R}^{3}\), \( \bm{y} := (\tilde{y}_1, \, \cdots, \, \tilde{y}_p) \in \mathbb{R}^m \),
and  \( \bm{u} := \Delta \in \R^6\),
the error dynamics \eqref{eq:observer_dynamics} with outputs \eqref{eq:observer_output} can be rewritten in the form \eqref{eq:riccati_dynamics},
with 
\begin{align*} 
    A(\tilde{\lambda}, t) &:= \begin{bmatrix}
        0_{3,3} & \hat{R} \\
        0_{3,3} & 0_{3,3}
    \end{bmatrix}, \, C(\tilde{\lambda}, t) := \begin{bmatrix}
        C_1(\tilde{\lambda}, t) & 0_{1,3}  \\
        \vdots & \vdots \\ C_p(\tilde{\lambda}, t) & 0_{1,3}
    \end{bmatrix},
\end{align*}
with $C_i(\tilde{\lambda}, t) := \Lambda_i^\top \hat{R}^\top b_i^\times$, $i=1, \dots,p$.
In light of \eqref{eq:riccati_input}, we choose the innovation $\Delta = -P C^\top Q \bm{y}$,
where $P(t)$ is the solution to the CRE \eqref{eq:CRE} associated with \eqref{eq:observer_dynamics}.
According to Proposition~\ref{proposition: uniform observability to stability}, the exponential stability of the equilibrium $\bm{x}=0$ is related to the uniform observability of the pair $(A^\star(t) := A(0,t), C^\star(t) := C(0,t))$, given by 
\begin{align} \label{eq:AC_star_biased}
    A^\star(t) &:= \begin{bmatrix}
        0_{3,3} & R \\
        0_{3,3} & 0_{3,3}
    \end{bmatrix}, & C^\star(t) &:= \begin{bmatrix}
        C_1^\star(t) & 0_{1,3}  \\
        \vdots & \vdots \\ C_p^\star(t) & 0_{1,3}
    \end{bmatrix},
\end{align}
with $C_i^\star(t) := \Lambda_i^\top R^\top b_i^\times$, $i=1, \dots,p$.
It therefore remains to provide explicit conditions under which uniform observability of $(A^\star(t), C^\star(t))$ is guaranteed. 

\subsection{Unbiased case} \label{sec:unbiased_case}
To simplify the analysis, we first consider the unbiased scenario, \textit{i.e.,} $d \equiv 0$. This allows us to separate the analysis and treat the simpler attitude-only system first, then extend to the full biased problem.
In this case, system \eqref{eq:observer_dynamics} reduces to the attitude error $\tilde{\lambda}$ alone:
\begin{subequations} \label{eq:system_dynamics_unbiased}
    \begin{align}
        \dot{\tilde{\lambda}} &= \Delta_R + O(|\tilde{\lambda}||\Delta_R|) + O(|\tilde{\lambda}|^2), \label{eq:state_equation_unbiased} \\
       \bm{y} &=
       C_\lambda(\tilde{\lambda},t) \tilde{\lambda} + O(|\tilde{\lambda}|^2),\label{eq:measurement_equation_unbiased}
    \end{align}
\end{subequations}
where $A_\lambda := 0$, and $ C_\lambda(\tilde{\lambda},t) := [C_1^\top, \dots, C_p^\top]^\top$. It follows directly that $A_\lambda^\star \equiv A_\lambda$ and $C_\lambda^\star(t) := [C_1^\star, \dots, C_p^\star]^\top$.

The following lemma is a direct application of Proposition~\ref{proposition: uniform observability to stability}, leveraging the definition of observability Gramian of the pair $(A_\lambda^\star,C_\lambda^\star(t))$ to establish local exponential stability of the attitude error $\tilde{R}$, which will be instrumental for the stability analysis of the full error $(\tilde{R}, \tilde{d})$.

\begin{lemma}\label{lemma:general_condition_lemma_1}
 Suppose that the input signal \(\Omega\) and the measurement directions $a_j^{(i)},b_i$ are continuous and bounded, and the observability Gramian
\begin{equation} \label{eq:condition_of_lemma1}
    \begin{aligned}
    W^{A_\lambda^\star,C_\lambda^\star}(t, t + \delta) &:=\! -\frac{1}{\delta}\!\int_t^{t+\delta}\!\sum_{i=1}^p b_i^\times R(s)\Lambda_i\Lambda_i^\top R^{\top} (s)\,b_i^\times\,ds,
    \end{aligned}    
\end{equation}
satisfies \eqref{eq:main Uniform Observability condition}.
Then, the equilibrium \(\tilde{R} = I_3\) is locally exponentially stable.
\end{lemma}

This result guarantees local convergence while relaxing the sufficient condition of Lemma~1 in \cite{alnahhal2025scalar}. In fact, the linear formulation in $\mathbb{R}^9$ imposes a stricter uniform observability requirement, whereas the intrinsic formulation on $\SO(3)$ requires uniform observability only in three dimensions. Consequently, this result strengthens Lemma~1 in \cite{alnahhal2025scalar} by establishing a necessary and sufficient uniform observability condition that can be applied across different scenarios.

To illustrate this, consider the classical and simplest case involving constant directions $a_j^{(i)}, b_i \in \mathbb{S}^2$. The next corollary provides explicit conditions for uniform observability of $(A_\lambda^\star,C_\lambda^\star(t))$ for the case of two and three scalar measurements ($m=2,3$) with two non-collinear inertial directions $b_1$ and $b_2$, which, \textit{e.g.}, might represent the inertial directions available from the pair accelerometer-magnetometer. Notice how $m=3$ is the minimum number of scalar measurements that guarantees uniform observability for $R$ constant.

\begin{definition} \label{def:vector_pe}
For any vector $\alpha(t) \in \mathbb{R}^3$, define
\begin{equation}
    U_{t,\delta}(\alpha) = \frac{1}{\delta} \int_{t}^{t+\delta} \alpha(s) \alpha^\top(s) \, ds,
\end{equation}
with eigenvalues $\lambda_1(t,\delta) \leq \lambda_2(t,\delta) \leq \lambda_3(t,\delta)$. Then $\alpha(t)$ is called \emph{strongly persistently exciting} if there exist $\delta, \beta > 0$ such that $\lambda_1(t,\delta) \geq \beta$ for all $t \geq 0$. It is \emph{weakly persistently exciting} if $\lambda_2(t,\delta) \geq \beta$. If $\mathrm{rank}(U_{t,\delta}(\alpha)) = 1$, then $\alpha(t)$ cannot be persistently exciting.
\end{definition}

\begin{corollary}\label{corollary:corollary_unbiased}
    Let $a_j^{(i)},b_i \in \mathbb{S}^2$ in~\eqref{eq:general_output} be constant known vectors. We distinguish two cases:
\begin{itemize}
        \item[1)] If $m=2$ with two non-collinear inertial directions $b_1, b_2$, i.e. $\Lambda_1 = a_1^{(1)},\Lambda_2=a_1^{(2)}$, then uniform observability is guaranteed if either:
        \begin{itemize}
            \item [i)] $R\Lambda_1$ and $R\Lambda_2$ are strongly persistently exciting.
            \item [ii)] Only $R\Lambda_1$ is strongly persistently exciting, and there exist $\delta,\mu>0$ such that 
            \begin{equation}\label{eq:corollary1_cond_case2}
                \frac{1}{\delta}\int_t^{t+\delta}\det([R(s)\Lambda_2\ b_1\  b_2])^2 ds >  \mu,
            \end{equation}
            or equivalently, only $R\Lambda_2$ is strongly persistently exciting and $R\Lambda_1$ satisfies \eqref{eq:corollary1_cond_case2}.
            \item[iii)] $R\Lambda_1$ and $R\Lambda_2$ are weakly persistently exciting and there exist $\delta,\mu>0$ such that for $i\in\{1,2\}$:
            \begin{equation}\label{eq:corollary1_cond2_case2}
                \frac{1}{\delta}\int_t^{t+\delta}\det([R(s)\Lambda_i\ b_1\  b_2])^2 ds >  \mu.
            \end{equation}
            \end{itemize}
        \item[2)] If $m = 3$ with two non-collinear inertial directions $b_1, b_2$ and $\Lambda_1=\left[a_1^{(1)}\ a_2^{(1)}\right], \Lambda_2=a_1^{(2)}$, then uniform observability is guaranteed if either:
        \begin{itemize}
            \item[i)] there exist $\delta,\mu>0$ such that
            \begin{subequations}
                \begin{align}
                    &\frac{1}{\delta}\int_t^{t+\delta}\det([R(s)\Lambda_1 \ b_1])^2ds > \mu, \label{eq:corollary_1_condition Ra1Ra2b1}\\
                    &\frac{1}{\delta}\int_t^{t+\delta}\det([R(s)\Lambda_2 \ b_1 \ b_2])^2 ds> \mu. \label{eq:corollary_1_condition Ra3b1b2}    
                \end{align}
            \end{subequations}
            \item[i$^\star$)] Assuming $R$ constant,
            \begin{equation*}
            \det([R\Lambda_1\ b_1]) \neq 0,\quad 
            \det([R \Lambda_2 \ b_1 \ b_2]) \neq 0.
            \end{equation*}
        \end{itemize}
    \end{itemize}
\end{corollary}
The proof, reported in Appendix~\ref{proof:corollary 1}, follows by direct inspection of the observability Gramian \eqref{eq:condition_of_lemma1}, which, for constant $a_j^{(i)}, b_i \in \mathbb{S}^2$, can be expressed as 
\begin{align*} 
    W^{A_\lambda^\star,C_\lambda^\star}(t, t + \delta) = - \sum_{i=1}^p b_i^\times \left(\sum_{j=1}^{n_i}U_{t,\delta}(Ra_j^{(i)})\right) b_i^\times.
\end{align*}

\subsection{Biased case}
We now extend the stability analysis to system \eqref{eq:observer_dynamics}, where the bias must also be estimated. The condition established in Lemma \ref{lemma:general_condition_lemma_1} is a prerequisite to guarantee the stability of the attitude subsystem \eqref{eq:system_dynamics_unbiased}. Building on this, we turn to the full error dynamics $\bm{x}$. The following lemma provides a sufficient condition for the uniform observability of $(A^\star(t), C^\star(t))$.

\begin{lemma}
    \label{lemma:general_condition_lemma_2}
Suppose that the input signal \(\Omega\) is continuous and bounded, that the measurement directions $a_j^{(i)}, b_i$ in~\eqref{eq:general_output} are uniformly continuous and bounded, and that condition~\eqref{eq:condition_of_lemma1} is verified. 
Moreover, assume that there exist constants \(\delta, \mu > 0\) such that, for all \(t \geq 0\), 
    \begin{align} \label{eq:condition_of_Lemma2}
      &W^{\bar{C}^\star}(t,t+\delta) \\ &\qquad := \frac{1}{\delta} \int_{t}^{t+\delta} 
        \sum_{i=1}^p \bar{C}_i^{\star\top}(s,t,\delta)\bar{C}_i^{\star}(s,t,\delta) ds \succeq \mu I_3, \notag
    \end{align}
where $\bar{C}_i^\star(s,t,\delta) := C_i^\star(s)\left(\int_t^s R(\tau)d\tau - \rho(t,\delta)\right)$, and
\begin{align*}
    &\rho(t,\delta) := \frac1\delta\left(W^{A_\lambda^\star,C_\lambda^\star}\right)^{-1}\int_{t}^{t+\delta}\sum_{i=1}^pC_i^{\star\top} C_i^{\star}\int_t^s R(\tau)d\tau ds.
    \end{align*}
Consequently, the pair $(A^\star(t),C^\star(t))$ is uniformly observable and $(\tilde{R},\tilde{d})=(I_3,0)$ is locally exponentially stable.
\end{lemma}

The proof, provided in Appendix~\ref{sec:proof_lemma2}, is based on a Schur complement argument that links conditions~\eqref{eq:condition_of_lemma1}–\eqref{eq:condition_of_Lemma2} to the uniform observability of $(A^\star(t), C^\star(t))$. The result then follows from Proposition~\ref{proposition: uniform observability to stability}.
In general, condition~\eqref{eq:condition_of_Lemma2} is difficult to verify for time-varying vectors $a_j^{(i)}, b_i$. However, when the vectors $a_j^{(i)}, b_i$ are constant and mild additional assumptions are imposed, the condition simplifies considerably. The following corollary gives \textit{sufficient conditions} in the case $m=p=2$ and $\Lambda_1=\Lambda_2$. All other cases can be understood as relaxations of the conditions in this corollary.

\begin{corollary}\label{corollary:general_condition_corollary_1}
    Suppose that the input signal \(\Omega\) is uniformly continuous and bounded. Let $m=p=2$, $a_j^{(i)}, b_i \in\mathbb{S}^2$ in ~\eqref{eq:general_output} be constant with $\Lambda_1 = \Lambda_2$, and assume $b_1, b_2$ non-collinear. Moreover, assume
    that Lemma \ref{lemma:general_condition_lemma_1} is fulfilled,
    and that $\Omega$ is strongly persistently exciting in the sense of Definition \ref{def:vector_pe}.
Then, the equilibrium $(\tilde{R},\tilde{d})=(I_3,0)$ is locally exponentially stable.        
\end{corollary}
The proof, reported in Appendix~\ref{sec:proof_corollary}, proceeds by contradiction and shows that strong persistence of excitation of $\Omega$ is a sufficient condition to guarantee that the derivative of the integrand of \eqref{eq:condition_of_Lemma2} is not zero.
\begin{remark}
    The scenario with two scalar measurements ($m=p=2$) with constant vectors $a_j^{(i)},b_i$, such that $\Lambda_1=\Lambda_2$, requires a stronger form of persistence of excitation for the biased case than for the unbiased one. In particular, 
    while Corollary~\ref{corollary:general_condition_corollary_1} requires strong persistence of excitation of $\Omega$, Corollary~\ref{corollary:corollary_unbiased} can be satisfied also by constant $\Omega$.
\end{remark}
\begin{remark}
In the case of measurements along constant inertial directions, the results of this section are consistent with the results in \cite{Mahony_Hamel_Pflimlin,batista2012sensor}, as also for scalar measurements, at least two non-collinear inertial directions are required for attitude reconstruction.
When dealing with time-varying inertial directions $b_i(t)$, even a single scalar measurement ($m=1$) can satisfy the persistent excitation conditions of Lemmas \ref{lemma:general_condition_lemma_1} and \ref{lemma:general_condition_lemma_2}; for instance, a Pitot tube with sufficiently varying airspeed provides the necessary excitation \cite{alnahhal2025scalar}.
\end{remark}

\section{Experimental Results}
This section presents experimental results to evaluate the performance of the proposed observer \eqref{eq:biased_obs_dyn}.
The evaluation is based on inertial data from the BROAD dataset \cite{laidig2021broad}, acquired using a 9-axis Myon aktos-t IMU at a sampling rate of $286$ Hz. This dataset is well-suited for benchmarking attitude estimation algorithms, as it provides accurate and well-calibrated inertial measurements from carefully executed motion sequences.
The IMU was mounted on a 3D-printed rigid body with reflective markers to ensure precise ground-truth orientation tracking.
The reference measurements in this setup are the gravity direction in the inertial frame $g_0 = e_3$, and the local magnetic field in the inertial frame $m_\circ$. Their corresponding body-frame measurements are provided by the accelerometer and magnetometer, respectively.
We use the BROAD dataset's first three sequences, A, B, and C, that consist of undisturbed slow pure rotations. These sequences serve two complementary purposes. First, they provide an ideal benchmark to validate the theoretical results from Section~\ref{sec:observer_design}, since their rotational motions are sufficiently varied to satisfy the persistent excitation condition stated in Corollary~\ref{corollary:general_condition_corollary_1}, which guarantees the theoretical convergence of the proposed design even when reduced to two scalar measurements. Second, they allow us to examine how the estimation accuracy evolves as the number of scalar measurements is progressively reduced.

The dataset's recorded IMU angular rate measurements contain a residual turn-on bias of $0.17 \mathrm{deg}$/s on average per axis \cite{laidig2021broad}. 
To assess the effect of gyro bias compensation, two observer implementations are considered: (i) the attitude-only observer outlined in Section \ref{sec:unbiased_case}, and (ii) the joint attitude-gyro bias observer \eqref{eq:biased_obs_dyn}.
Both observers are evaluated under four different sets of scalar measurements obtained by selecting different combinations of accelerometer and magnetometer axes, starting from full 3-axis measurements for both sensors down to only a single axis from each sensor. The measurement sets used in each configuration are summarized in Table~\ref{tab:sensor_components}.
\begin{table}[h!]
\centering
\caption{Accelerometer and magnetometer components used in different measurement configurations.}
\begin{tabular}{c c c c c}
\hline
\textbf{Scalars} & \textbf{Acc $\Lambda_1$} & \textbf{Mag $\Lambda_2$} & \textbf{Removed} \\
\hline
Six (all) & $\begin{bmatrix}
   e_1\;e_2\;e_3 
\end{bmatrix}$ & $\begin{bmatrix}
   e_1\;e_2\;e_3 
\end{bmatrix}$ & --- \\
Four & $\begin{bmatrix}
   e_2\;e_3 
\end{bmatrix}$ & $\begin{bmatrix}
   e_1\;e_2 
\end{bmatrix}$ & Acc: $e_1$, Mag: $e_3$ \\
Three & $\begin{bmatrix}
   e_2\;e_3 
\end{bmatrix}$ & $e_2$ & Acc: $e_1$, Mag: $e_1,e_3$ \\
Two & $e_2$ & $e_2$ & Acc: $e_1,e_3$, Mag: $e_1,e_3$ \\
\hline
\end{tabular}
\label{tab:sensor_components}
\end{table}

For completeness, we also report results obtained with the complementary filter \cite{Mahony_Hamel_Pflimlin}, commonly used as a baseline in attitude estimation. However, it should be noted that a direct performance comparison with this filter is not strictly meaningful due to its fundamentally different design. 

The observers were initialized with $\hat{R}(0)=I_3$ and $\hat{d}(0)=0$, with $P(0)=0.5 I_3$ and $V=0.005 I_3$ for the attitude-only observer, and $P(0)=0.5 I_6$ and $V=0.005 I_6$ for the full attitude+gyro bias observer, and $Q=0.05 I_m$ for both.
For the complementary filter, the attitude innovation gain was set to $2.5$ and the bias adaptation gain to $0.1$.

\subsection{Results and discussion}

The attitude estimates obtained using the complementary filter and the proposed observer are compared against the Optitrack motion capture ground truth from the BROAD dataset. Their performance is evaluated using the root mean square error (RMSE)
on the angular distance on $\SO(3)$, defined by $\theta := \arccos(\frac{1}{2}(\mathrm{tr}(\tilde{R}) - 1))$.
The results are summarized in Table~\ref{tab:rmse_results}.

\begin{table*}[t]
\centering
\caption{RMSE (in degrees) of the attitude estimation error $\theta$ 
for the complementary filter and the proposed observers on Sequences A, B, and C of the BROAD dataset. 
For each sequence, results are reported with and without bias compensation.}
\label{tab:rmse_results}

\setlength{\tabcolsep}{5pt}
\renewcommand{\arraystretch}{1.05}

\begin{tabular}{|l|cc|cc|cc|}
\multicolumn{1}{c}{} & \multicolumn{2}{c}{\textbf{Sequence A}}
 & \multicolumn{2}{c}{\textbf{Sequence B}}
 & \multicolumn{2}{c}{\textbf{Sequence C}} \\
\toprule
\textbf{Observer}
 & Attitude only & Bias compensation
 & Attitude only & Bias compensation
 & Attitude only & Bias compensation \\
\midrule

Complementary \cite{Mahony_Hamel_Pflimlin}
& \textbf{1.660} & \textbf{1.701} & \textbf{1.503} & 1.480 & \textbf{4.098} & 3.885 \\

Proposed -- 6 scalars
& 2.723 & 1.903 & 1.829 & \textbf{1.295} & 4.449 & \textbf{3.529} \\

Proposed -- 4 scalars
& 2.677 & 2.087 & 3.499 & 1.770 & 5.634 & 3.621 \\

Proposed -- 3 scalars
& 3.037 & 2.399 & 4.347 & 2.552 & 5.443 & 3.563 \\

Proposed -- 2 scalars
& 3.502 & 2.835 & 5.279 & 3.242 & 5.543 & 3.665 \\

\bottomrule
\end{tabular}
\end{table*}

Overall, the Riccati observer with full vector measurements (6 scalars) achieves the lowest total RMSE and consistently accurate estimates across all sequences, which indeed demonstrates the benefit of exploiting the complete measurement set. 
The complementary filter (included as a baseline) yields comparable RMSE results.
Reducing the number of scalar measurements naturally leads to a slight increase in attitude errors. Still, even the two-scalar measurements observer achieves low RMSEs, demonstrating the effectiveness of the design under reduced measurement conditions under sufficiently persistently exciting motion.

\begin{figure}[t]
    \centering    \includegraphics[width=1.\linewidth, trim=1.9cm 0cm 1.8cm 0cm, clip]{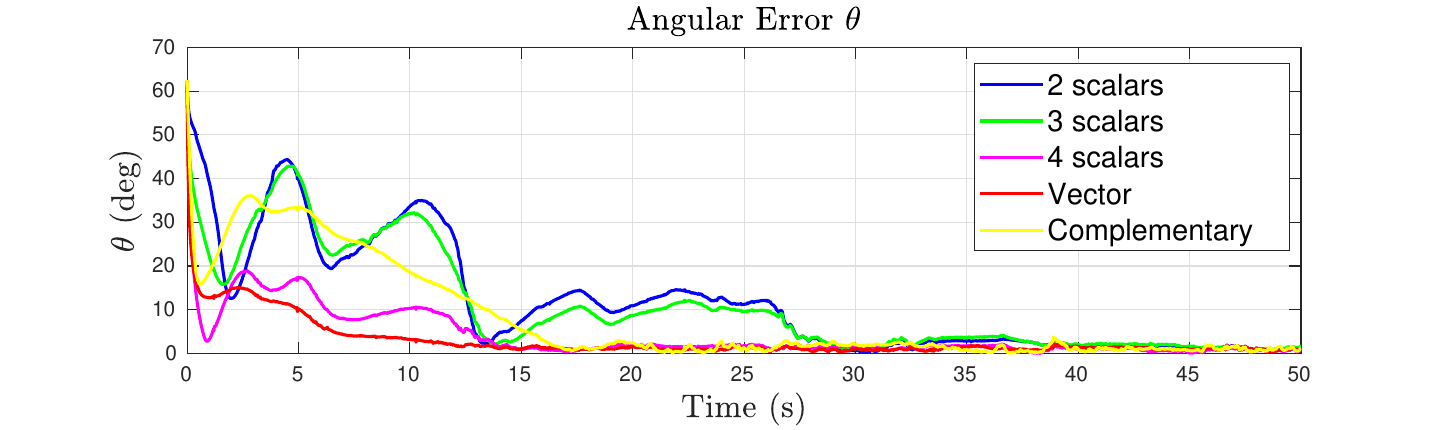}
    \caption{Evolution of the attitude error $\theta$ for all observers with initial condition $\hat{R}(0)$ set to a $45\mathrm{deg}$ rotation about each axis.}
    \label{fig:attitude_errors}
\end{figure}

An additional experiment was performed on sequence B to illustrate the convergence transients of observer \eqref{eq:biased_obs_dyn}. The attitude estimate was initialized with $\hat{R}(0)$ corresponding to a rotation of $45\mathrm{deg}$ about each axis. The resulting error convergence is shown in Fig.~\ref{fig:attitude_errors}. 
The results confirm that all observers converge to the true attitude.
Convergence is fastest with full 3-axis vector measurements, followed by 4 scalars, while the 3- and 2-scalar configurations are slightly slower. 
This is directly related to the amount of excitation provided by the selected measurements in the conditions of Lemmas \ref{lemma:general_condition_lemma_1} and \ref{lemma:general_condition_lemma_2}.

\section{Conclusion}
We proposed a deterministic Riccati-based observer for attitude estimation on $\SO(3)$ using scalar measurements, explicitly accounting for gyroscope bias. 
The framework avoids high-dimensional embeddings and establishes persistence-of-excitation conditions linking uniform observability of the linearized dynamics to local exponential stability. 
A key theoretical result is that two scalar measurements under suitable excitation suffice for attitude observability, and that three are enough in the static case, matching the intrinsic dimension of $\SO(3)$. 
Experiments on the BROAD dataset confirmed these findings: the observer retained stable convergence even when reduced to two scalars, with accuracy degradation following the expected trend as measurements were progressively removed. 
This demonstrates the robustness and practical relevance of scalar-based estimation frameworks, especially in scenarios where full vector information is unavailable. 
Future work will focus on enlarging the domain of convergence of the proposed estimation scheme and on exploring lightweight constant-gain designs.

\section{Appendix}
\subsection{Proof of Corollary~\ref{corollary:corollary_unbiased}}\label{proof:corollary 1}
Since $a_j^{(i)}, b_i \in \mathbb{S}^2$ are constant, the observability Gramian can be expressed as 
\begin{equation} 
    W^{A_\lambda^\star,C_\lambda^\star}(t, t + \delta) = - \sum_{i=1}^p b_i^\times \left(\sum_{j=1}^{n_i}U_{t,\delta}(Ra_j^{(i)})\right) b_i^\times.
\end{equation}
The uniform observability condition in \eqref{eq:condition_of_lemma1} is equivalent to $v^\top W^{A_\lambda^\star,C_\lambda^\star} v \geq \mu > 0$ for any vector $v \in \mathbb{S}^2$, which can be rewritten as
\begin{equation} \label{eq:cond_corollary_1}
    \sum_{i=1}^p \bar{v}_i^\top U_{t,\delta}\left(\sum_{j=1}^{n_i}(Ra_j^{(i)})\right) \bar{v}_i \geq \mu, \quad \bar{v}_i = b_i \times v.
\end{equation}
It is straightforward that if $m=1$ then choosing $v = b_1$ yields $\bar{v}_1 = 0$, which violates \eqref{eq:cond_corollary_1} and therefore $(A_\lambda^\star,C_\lambda^\star(t))$ is not uniformly observable. 

Case 1) For $m=2$, 
we get
$$ v^\top W^{A_\lambda^\star,C_\lambda^\star} v = \bar{v}_1^\top U_{t,\delta}(R a_1^{(1)}) \bar{v}_1 + \bar{v}_2^\top U_{t,\delta}(R a_1^{(2)}) \bar{v}_2. $$
It is clear that uniform observability cannot be guaranteed if $R$ is constant, since there exists $v$ which cancels both terms, given directly by $v = (R a_1^{(1)} \times b_1)\times (R a_1^{(2)} \times b_2)$.

(i) If $Ra_1^{(1)}$ and $Ra_1^{(2)}$ are strongly persistently exciting, \textit{i.e.,} $U_{t,\delta}(Ra_1^{(i)}) \succeq \lambda_{i1}(t, \delta) I_3 \succeq \beta I_3$, then $ \bar{v}_1^\top U_{t,\delta}(Ra_1^{(i)}) \bar{v}_1 \geq \beta |\bar{v}_i|^2$, $i=1,2$.
Moreover, since $b_1$ and $b_2$ are non-collinear, at least one of $\bar{v}_1, \bar{v}_2$ is non-zero for any $v$. 
Therefore, $$v^\top W^{A_\lambda^\star,C_\lambda^\star} v \geq \beta |\bar{v}_1|^2 + \beta |\bar{v}_2|^2 > 0.$$ 
It follows that $(A_\lambda^\star,C_\lambda^\star(t))$ is uniformly observable.

(ii) Assume that at least one $R a_1^{(i)}$, $i=1,2$ is strongly persistently exciting, \textit{i.e.,} $U_{t, \delta}(R a_1^{(i)}) \succeq \beta I_3$, then we directly obtain $v^\top W v \geq \beta |\bar{v}_i|^2$ for all $v \neq \pm b_i$ (\textit{i.e.,} $\bar{v}_i \neq 0$). If $v =\pm b_i$, then $v^\top W v = (b_l \times b_i)^\top U_{t, \delta}(Ra_1^{(l)}) (b_l \times b_i) $ for $l\in\{1,2\}$, $l \neq i$, which is strictly positive from \eqref{eq:corollary1_cond_case2}, due to the equivalence between triple product and determinant, \textit{i.e.,} $\det([Ra_1^{(i)}\ b_1\ b_2]) = a_1^{(i)\top} R^\top b_1^\times b_2$.

(iii) Assume $Ra_1^{(1)}$ and $Ra_1^{(2)}$ are both weakly persistently exciting, \textit{i.e.,} 
$\mathrm{rank}(U_{t,\delta}(R a_1^{(i)}))=2$, $i=1,2$. 
This is only possible if there exists $\bar t\geq 0$ such that $R(t)a_1^{(1)}$ and $R(t)a_1^{(2)}$ remain on the same plane for all $t \geq \bar t$. Let $\eta = a_1^{(1)} \times a_1^{(2)}$ denote the direction orthogonal to $\mathrm{span}\{a_1^{(1)},a_1^{(2)}\}$, that is, $ \mathrm{ker}(U_{t,\delta}(R a_1^{(1)}))=\mathrm{ker}(U_{t,\delta}(R a_1^{(2)}))=\mathrm{span}\{R\eta\}$ for all $t\geq \bar t$. 
If $v = \pm b_i$, then $v^\top W v = (b_l \times b_i)^\top U_{t, \delta}(Ra_1^{(l)}) (b_l \times b_i) $ for $l\in\{1,2\}$, $l \neq i$, then $v^\top W v > 0$ is implied by \eqref{eq:corollary1_cond2_case2}.
If $v \neq \pm b_1, \pm b_2$, then $\bar{v}_1 \neq 0$ and $\bar{v}_2 \neq 0$, and the only way for both terms to vanish simultaneously for $t\geq \bar t$ is if both $\bar{v}_1$ and $\bar{v}_2$ are parallel to $R\eta$. This is only possible when both $b_1$ and $b_2$ lie in the plane of $R a_1^{(1)}$ and $R a_1^{(2)}$, which is prevented by \eqref{eq:corollary1_cond2_case2}.
It follows that $(A_\lambda^\star,C_\lambda^\star(t))$ is uniformly observable.

Case 2) For $m=3$, assume two non-collinear inertial directions $b_1, b_2$ and $\Lambda_1=\left[a_1^{(1)}\ a_2^{(1)}\right], \Lambda_2=a_1^{(2)}$. Let $\bar{v}_1 = b_1 \times v$ and $\bar{v}_2 = b_2 \times v$, then for any $v \in \mathbb{S}^2$,
\begin{align*}
v^\top &W v \\&= \bar{v}_1^\top ( U_{t,\delta}(Ra_1^{(1)}) + U_{t,\delta}(Ra_2^{(1)}) ) \bar{v}_1 
+ \bar{v}_2^\top U_{t,\delta}(Ra_1^{(2)}) \bar{v}_2.    
\end{align*}

(i) Condition \eqref{eq:corollary_1_condition Ra1Ra2b1} guarantees that $\bar{v}_1^\top ( U_{t,\delta}(Ra_1^{(1)}) + U_{t,\delta}(Ra_2^{(1)}) ) \bar{v}_1 = 0$ only if $v =\pm b_1$. If $v = \pm b_1$, the term $\bar{v}_1^\top (U_{t,\delta}(Ra_1^{(1)}) + U_{t,\delta}(Ra_2^{(1)})) \bar{v}_1$ vanishes. In this case,
$$ v^\top W v = (b_2 \times b_1)^\top U_{t,\delta}(Ra_1^{(2)}) (b_2 \times b_1),$$
which is strictly positive from \eqref{eq:corollary_1_condition Ra3b1b2}.

(i$^\star$) To ensure that $\bar{v}_1^\top ( U_{t,\delta}(Ra_1^{(1)}) + U_{t,\delta}(Ra_2^{(1)}) ) \bar{v}_1 > 0$ for all $v \neq \pm b_1$, the three vectors $Ra_1^{(1)}$, $Ra_2^{(1)}$ and $b_1$ must be linearly independent. This is satisfied if 
\begin{align*}
    \det([Ra_1^{(1)} \ Ra_2^{(1)} \ b_1])=\det([R\Lambda_1\ b_1]) \neq 0.
\end{align*}
If $v = \pm b_1$, then the term $\bar{v}_1^\top (U_{t,\delta}(Ra_1^{(1)}) + U_{t,\delta}(Ra_2^{(1)})) \bar{v}_1$ vanishes. In this case, 

$$ v^\top W v = (b_2 \times b_1)^\top U_{t,\delta}(Ra_1^{(2)}) (b_2 \times b_1).$$
This is strictly positive if the vectors $R a_1^{(2)}$, $b_1$, and $b_2$ form a non-degenerate volume in $\mathbb{R}^3$, \textit{i.e.,} $\det([R\Lambda_2 \ b_1 \ b_2]) \neq 0$.

\subsection{Proof of Lemma~\ref{lemma:general_condition_lemma_2}}
\label{sec:proof_lemma2}
    By Peano-Baker series expansion, the state transition matrix associated with \eqref{eq:AC_star_biased} is
    \begin{equation*}
        \Phi(s,t) = \begin{bmatrix}
            I_3 & \int_t^s R(\tau)d\tau\\
            0_{3\times3} & I_3
        \end{bmatrix}.
    \end{equation*}
    Hence, the observability Gramian associated with \eqref{eq:AC_star_biased} is
    \begin{align*}
        &W^{A^\star,C^\star}(t,t+\delta)= \begin{bmatrix}
            W^{A_\lambda^\star,C_\lambda^\star}(t,t+\delta) & M(t,\delta)\\
            M^\top(t,\delta) & H(t,\delta)
        \end{bmatrix}
        \label{eq:gramian_full}
    \end{align*}
    where 
    \begin{align*}
        M &:=\frac1\delta\int_t^{t+\delta}\sum_{i=1}^pC_i^{\star\top}C_i^\star\Bigl(\int_t^s R(\tau)d\tau\Bigr) ds,\\
H &:=\frac1\delta\int_t^{t+\delta}\Bigl(\int_t^s R^\top(\tau)d\tau\Bigr)
\Bigl(\sum_{i=1}^p C_i^{\star\top}C_i^\star\Bigr)\Bigl(\int_t^s R(\tau)d\tau\Bigr)ds.
    \end{align*}
    Since condition~\eqref{eq:condition_of_lemma1} holds, the matrix $W^{A_\lambda^\star,C_\lambda^\star}(t,t+\delta)$ is uniformly positive definite and therefore invertible for all \(t\geq 0\). We may thus define the Schur complement of $W^{A^\star,C^\star}(t,t+\delta)$ with respect to $W^{A_\lambda^\star,C_\lambda^\star}(t,t+\delta)$ by
\begin{equation}\label{eq:schur_complement}
W^{A^\star,C^\star}/W^{A_\lambda^\star,C_\lambda^\star}(t,t+\delta) := H - M^\top \left(W^{A_\lambda^\star,C_\lambda^\star}\right)^{-1} M.
\end{equation}
It is readily verifiable that \eqref{eq:schur_complement} and the left-hand side of \eqref{eq:condition_of_Lemma2} are equivalent. Indeed, notice that
\begin{align*}
    \sum_{i=1}^p \bar{C}_i^{\star} =
    C_\lambda^\star\left(\int_t^s R d\tau - \left(W^{A_\lambda^\star,C_\lambda^\star}\right)^{-1} M\right).
\end{align*}
Then, from \eqref{eq:condition_of_Lemma2}, it follows that
\begin{align*}
    &W^{\bar{C}^\star}(t,t+\delta) =\\&= \frac{1}{\delta} \int_{t}^{t+\delta} 
        \sum_{i=1}^p \bar{C}_i^{\star\top}(s,t,\delta)\bar{C}_i^{\star}(s,t,\delta) ds = \\
        &= H - \frac{1}{\delta}\int_{t}^{t+\delta} 
        \int_t^s R^\top d\tau C_\lambda^{\star\top}C_\lambda^\star ds \left(W^{A_\lambda^\star,C_\lambda^\star}\right)^{-1} M \\
        &\;- M^\top \left(W^{A_\lambda^\star,C_\lambda^\star}\right)^{-1} \frac{1}{\delta}\int_{t}^{t+\delta} 
        C_\lambda^{\star\top}C_\lambda^\star\int_t^s R d\tau ds\\
        & \;+  M^\top \left(W^{A_\lambda^\star,C_\lambda^\star}\right)^{-1}\frac{1}{\delta} \int_{t}^{t+\delta} C_\lambda^{\star\top} C_\lambda^{\star} ds\left(W^{A_\lambda^\star,C_\lambda^\star}\right)^{-1} M\\
        &=H - M^\top \left(W^{A_\lambda^\star,C_\lambda^\star}\right)^{-1} M =W^{A^\star,C^\star}/W^{A_\lambda^\star,C_\lambda^\star}.
\end{align*}

By the boundedness assumption on $a_i,b_i$ and $\Omega$, by the Gershgorin circle theorem, it follows that the maximum eigenvalues of $W^{A^\star,C^\star},\ W^{A_\lambda^\star,C_\lambda^\star}$ and $W^{A^\star,C^\star}/W^{A_\lambda^\star,C_\lambda^\star}$ are upper bounded by a constant value. Moreover, conditions \eqref{eq:condition_of_lemma1} and \eqref{eq:condition_of_Lemma2} are equivalent to the existence of $\mu>0$ such that $\lambda_{min}\left(W^{A_\lambda^\star,C_\lambda^\star}\right)>\mu$ and $\lambda_{min}\left(W^{A^\star,C^\star}/W^{A_\lambda^\star,C_\lambda^\star}\right)>\mu$, where $\lambda_{min}(\cdot)$ denotes the minimum eigenvalue of a matrix. Since the Schur complement satisfies the determinant property
    \begin{align*}\label{eq:Schur_determinant}
        \text{det}\left(W^{A^\star,C^\star}\right) = \text{det}\left(W^{A_\lambda^\star,C_\lambda^\star}\right)\text{det}\left(W^{A^\star,C^\star}/W^{A_\lambda^\star,C_\lambda^\star}\right),
    \end{align*}
    it follows that there exists $\mu^\star>0$ such that $\lambda_{min}\left(W^{A^\star,C^\star}\right)>\mu^\star$, which is an equivalent definition for uniform observability of $(A^\star, C^\star)$.
     
The local exponential stability of $(\tilde R,\tilde d)=(I_3,0)$ then follows from Proposition~\ref{proposition: uniform observability to stability}.

\subsection{Proof of Corollary~\ref{corollary:general_condition_corollary_1} }
\label{sec:proof_corollary}

Pick $\delta>0$ large enough to satisfiy both the strong persistence of excitation of $\Omega$, both the conditions of Lemma 1.

Let us proceed by contradiction, namely suppose that $W^{A^\star,C^\star}$ is not uniformly positive definite. Then there exist a sequence $\{t_k\}_{k\in\mathbb{N}}$ and $v=(v_1,v_2)\in\mathbb{S}^5$, with $v_1,v_2\in\R^3$, such that
\begin{equation}\label{eq:lim contradiction gramian}
    \lim_{k\to\infty}\frac{1}{\delta}\int_{0}^{\delta}\norm{C_\lambda^\star (t_k+s)\left(v_1 + \int_{0}^sR(t_k+\tau)v_2\,d\tau\right)}^2\,ds = 0,
\end{equation}

If $v_2=0$, then we reach a contradiction right away due to Lemma 1. Therefore, in the following we will assume $v_2\neq 0$.

By the Arzelà-Ascoli theorem, since $\Omega$ is bounded and uniformly continuous, there exist $\bar R(s)\in \SO(3)$ and $\bar\Omega(s)\in\R^3$ such that
\begin{align*}
    \lim_{k\to\infty} R(t_k+s) = \bar R(s),\quad \lim_{k\to\infty} \Omega(t_k+s) = \bar \Omega(s),
\end{align*}
uniformly on $s\in[0,\delta]$, with $\dot{\bar R}=\bar R\bar\Omega^\times$. Notice that, $\bar \Omega$ is strongly persistently exciting, by strong persistence of excitation of $\Omega$, and that by the assumptions of Lemma 1, there exists $\mu>0$ such that, for all $\eta\in\mathbb{S}^2$,
\begin{align}\label{eq:limit_lemma1}
\frac1\delta\int_0^\delta\sum_{i=1}^2\left|\Lambda^\top \bar R^\top(s)b_i^\times \eta\right|^2ds \geq \mu .
\end{align}

Denote
\begin{align*}
    z(s):=v_1 + \int_{0}^s\bar R(\tau)v_2\,d\tau.
\end{align*}
Then, \eqref{eq:lim contradiction gramian} implies that for all $s\in[0,\delta]$,
\begin{equation}\label{eq:zero inside integral}
    \Lambda^\top \bar R^\top(s)b_i^\times z(s) = 0,\quad i=1,2.
\end{equation}
Denote,
\begin{align*}
    d(s):= \det[\bar R(s)\Lambda\ \ b_1\ \ b_2],
\end{align*}
and consider the set of $s\in[0,\delta]$ such that $d(s) \neq 0$. Then \eqref{eq:zero inside integral} implies
\begin{align*}
    z(s)\in \Span\{\bar R(s)\Lambda,b_1\},
\quad
z(s)\in \Span\{\bar R(s)\Lambda,b_2\}.
\end{align*}
Since $\bar R(s) \Lambda,\ b_1$ and $b_2$ are linearly independent, it follows that there exists $\alpha(s)\in\R$ such that
\begin{align*}
z(s) = \alpha(s) \bar R(s)\Lambda.
\end{align*}
Differentiating this expression gives
\begin{align}\label{eq:Rv2 = derivative}
    \bar R(s) v_2 = \dot \alpha(s) \bar R(s) \Lambda + \alpha(s) \bar R(s)\bar\Omega^\times(s)\Lambda,
\end{align}
Denote $\Pi_\Lambda := I_3 - \Lambda\Lambda^\top$. Then, multiplying by $\bar R^\top$ and projecting onto $\Lambda^\perp$, one obtains
\begin{align}\label{eq:OmegaLambda perp Piv2}
    \Pi_{\Lambda} v_2 = \alpha(s)\bar\Omega^\times(s)\Lambda.
\end{align}

Consider now the set where $d(s)=0$. Then, $\bar R(s)\Lambda \in \Span\{b_1,b_2\}$, hence there exist $\beta_1(s),\beta_2(s)\in\R$ such that
\begin{align*}
    \bar R(s)\Lambda = \beta_1(s)b_1+\beta_2(s) b_2.
\end{align*}
Substituting into \eqref{eq:zero inside integral} we get
\begin{align*}
    \Lambda^\top \bar R^\top(s)b_i^\times z(s) = \beta_{3-i}(s)b_{3-i}^\top (b_i^\times z(s)),\quad i=1,2,
\end{align*}
which imply
\begin{align}\label{eq:b1b2 z =0}
    (b_1\times b_2)^\top z(s) = 0.
\end{align}
Moreover, $d(s) =0$ is equivalent to
\begin{align}\label{eq:b1b2 RLambda =0}
    (b_1\times b_2)^\top \bar R(s) \Lambda = 0.
\end{align}
Since $\bar R(s)$ and $z(s)$ are absolutely continuous, the derivatives of \eqref{eq:b1b2 z =0} and \eqref{eq:b1b2 RLambda =0} are almost everywhere zero in the set where $d(s) = 0$, namely
\begin{align}\label{eq:b1b2 Rv2 = 0}
    (b_1\times b_2)^\top \bar R(s) v_2 = 0,\\
    (b_1\times b_2)^\top \bar R(s) \bar \Omega^\times(s)\Lambda = 0. \label{eq:b1b2 ROmLambda = 0}
\end{align}
Equations \eqref{eq:b1b2 RLambda =0}-\eqref{eq:b1b2 Rv2 = 0}-\eqref{eq:b1b2 ROmLambda = 0} imply that
\begin{align*}
    \Lambda,v_2,\bar\Omega^\times(s)\Lambda \perp \bar R^\top(s)(b_1\times b_2).
\end{align*}
Since $\Pi_\Lambda v_2$ and $\bar \Omega^\times(s)\Lambda$ are orthogonal to both $\Lambda$ and $\bar R^\top(s)(b_1\times b_2)$, then there exists a, possibly different, $\alpha(s)\in\R$, such that \eqref{eq:OmegaLambda perp Piv2} is satisfied also for $d(s) = 0$.

Assume now $\Pi_\Lambda v_2 \neq 0$. This means that for almost every $s\in[0,\delta]$, 
\begin{align*}
    \bar \Omega(s)\in \Span\{\Lambda,\Lambda^\times \Pi_{\Lambda}v_2\},
\end{align*}
or equivalently, that there exists $\xi\in\R^3$ such that for almost all $s\in[0,\delta]$
\begin{align*}
    \xi^\top\bar\Omega(s) = 0,
\end{align*}
which violates strong persistence of excitation of $\bar \Omega$. As a consequence, we can conclude that
\begin{align*}
    \Pi_\Lambda v_2 = 0,
\end{align*}
or equivalently, that there exists $\gamma\neq 0$, since $\norm{v_2}\neq 0$ and $\norm{\Lambda}=1$, such that
\begin{align*}
    v_2 = \gamma \Lambda.
\end{align*}
Substituting into \eqref{eq:Rv2 = derivative}, we obtain
\begin{align*}
    \dot\alpha(s) = \gamma,
\end{align*}
hence $\alpha(s)$ cannot be identically zero on any connected subset of $[0,\delta]$ of non-zero measure. On the set where $d(s) \neq 0$, \eqref{eq:OmegaLambda perp Piv2} reduces to
\begin{align*}
    \alpha(s)\bar\Omega^\times(s)\Lambda=0.
\end{align*}
We can conclude that
\begin{align*}
    \bar\Omega^\times(s)\Lambda=0,
\end{align*}
almost everywhere on every connected subset of $[0,\delta]$ where $d(s) \neq 0$. Consequently, on these sets
\begin{align*}
    \frac{d}{ds} \bar R(s) \Lambda = \bar R(s) \bar \Omega^\times(s)\Lambda=0,
\end{align*}
\textit{i.e.}, $\bar R(s)\Lambda$ is constant almost everywhere on every connected subset of $[0,\delta]$ where $d(s) \neq 0$. By continuity, this implies that $\bar R(s)\Lambda$ is constant on $(0,\delta)$. However, this condition contradicts \eqref{eq:limit_lemma1}, concluding uniform observability of the pair $(A^\star,C^\star)$. Local exponential stability follows from Proposition 1.

\bibliographystyle{IEEEtran} 
\bibliography{references} 

@article{markley1988attitude,
  author    = {F. Landis Markley},
  title     = {Attitude determination using vector observations and the singular value decomposition},
  journal   = {The Journal of the Astronautical Sciences},
  volume    = {36},
  number    = {3},
  pages     = {245--258},
  year      = {1988},
}

@article{berkane2017design,
  title={On the Design of Attitude Complementary Filters on {$SO(3)$}},
  author={Berkane, Soulaimane and Tayebi, Abdelhamid},
  journal={IEEE Transactions on Automatic Control},
  volume={63},
  number={3},
  pages={880--887},
  year={2017},
  publisher={IEEE}
}

@article{shuster1981three,
  title={Three-axis attitude determination from vector observations},
  author={Shuster, Malcolm David and Oh, S D\_},
  journal={Journal of guidance and Control},
  volume={4},
  number={1},
  pages={70--77},
  year={1981}
}

@article{wahba1965least,
  title={A least squares estimate of satellite attitude},
  author={Wahba, Grace},
  journal={SIAM review},
  volume={7},
  number={3},
  pages={409--409},
  year={1965},
  publisher={Society for Industrial and Applied Mathematics}
}

@article{hamel2018riccati,
   author = {Tarek Hamel and Claude Samson},
   doi = {10.1109/TAC.2017.2726179},
   issn = {0018-9286},
   issue = {3},
   journal = {IEEE Transactions on Automatic Control},
   month = {3},
   pages = {726-741},
   title = {Riccati Observers for the Nonstationary PnP Problem},
   volume = {63},
   year = {2018}
}

@article{Mahony_Hamel_Pflimlin,
   author = {Robert Mahony and Tarek Hamel and Jean-Michel Pflimlin},
   doi = {10.1109/TAC.2008.923738},
   isbn = {2008.923738},
   issn = {0018-9286},
   issue = {5},
   journal = {IEEE Transactions on Automatic Control},
   month = {6},
   pages = {1203-1218},
   title = {Nonlinear Complementary Filters on the Special Orthogonal Group},
   volume = {53},
   year = {2008}
}

@article{batista2012sensor,
   author = {P. Batista and C. Silvestre and P. Oliveira},
   doi = {10.1109/TAC.2012.2187142},
   issn = {0018-9286},
   issue = {8},
   journal = {IEEE Transactions on Automatic Control},
   month = {8},
   pages = {2095-2100},
   title = {Sensor-Based Globally Asymptotically Stable Filters for Attitude Estimation: Analysis, Design, and Performance Evaluation},
   volume = {57},
   year = {2012}
}

@article{zlotnik2016nonlinear,
   author = {David Evan Zlotnik and James Richard Forbes},
   doi = {10.1109/TAC.2016.2547222},
   issn = {0018-9286},
   issue = {1},
   journal = {IEEE Transactions on Automatic Control},
   month = {1},
   pages = {149-160},
   title = {Nonlinear Estimator Design on the Special Orthogonal Group Using Vector Measurements Directly},
   volume = {62},
   year = {2017}
}

@article{tayebi2006attitude,
   author = {A. Tayebi and S. McGilvray},
   doi = {10.1109/TCST.2006.872519},
   issn = {1063-6536},
   issue = {3},
   journal = {IEEE Transactions on Control Systems Technology},
   month = {5},
   pages = {562-571},
   title = {Attitude stabilization of a VTOL quadrotor aircraft},
   volume = {14},
   year = {2006}
}

@article{izadi2014rigid,
   author = {Maziar Izadi and Amit K. Sanyal},
   doi = {10.1016/j.automatica.2014.08.010},
   issn = {00051098},
   issue = {10},
   journal = {Automatica},
   month = {10},
   pages = {2570-2577},
   title = {Rigid body attitude estimation based on the Lagrange–d’Alembert principle},
   volume = {50},
   year = {2014}
}

@article{crassidis2007survey,
   author = {John L. Crassidis and F. Landis Markley and Yang Cheng},
   doi = {10.2514/1.22452},
   issn = {0731-5090},
   issue = {1},
   journal = {Journal of Guidance, Control, and Dynamics},
   month = {1},
   pages = {12-28},
   title = {Survey of Nonlinear Attitude Estimation Methods},
   volume = {30},
   year = {2007}
}

@article{lefferts1982kalman,
   author = {E.J. Lefferts and F.L. Markley and M.D. Shuster},
   doi = {10.2514/3.56190},
   issn = {0731-5090},
   issue = {5},
   journal = {Journal of Guidance, Control, and Dynamics},
   month = {9},
   pages = {417-429},
   title = {Kalman Filtering for Spacecraft Attitude Estimation},
   volume = {5},
   year = {1982}
}

@article{alnahhal2025scalar,
   author = {Hassan Alnahhal and Sifeddine Benahmed and Soulaimane Berkane and Tarek Hamel},
   doi = {10.1109/LCSYS.2025.3578470},
   issn = {2475-1456},
   journal = {IEEE Control Systems Letters},
   pages = {1862-1867},
   title = {Attitude Estimation Using Scalar Measurements},
   volume = {9},
   year = {2025}
}

@article{barrau2018invariant,
   author = {Axel Barrau and Silvère Bonnabel},
   doi = {10.1146/annurev-control-060117-105010},
   issn = {2573-5144},
   issue = {1},
   journal = {Annual Review of Control, Robotics, and Autonomous Systems},
   month = {5},
   pages = {237-257},
   title = {Invariant Kalman Filtering},
   volume = {1},
   year = {2018}
}

@article{hua2014implementation,
   author = {Minh-Duc Hua and Guillaume Ducard and Tarek Hamel and Robert Mahony and Konrad Rudin},
   doi = {10.1109/TCST.2013.2251635},
   issn = {1063-6536},
   journal = {IEEE Transactions on Control Systems Technology},
   month = {1},
   pages = {201-213},
   title = {Implementation of a Nonlinear Attitude Estimator for Aerial Robotic Vehicles},
   volume = {22},
   year = {2014}
}

@article{aeyels1998asymptotic,
   author = {D. Aeyels and R. Sepulchre and J. Peuteman},
   doi = {10.1007/BF02741883},
   issn = {0932-4194},
   issue = {1},
   journal = {Mathematics of Control, Signals, and Systems},
   month = {3},
   pages = {1-27},
   title = {Asymptotic stability for time-variant systems and observability: Uniform and nonuniform criteria},
   volume = {11},
   year = {1998}
}

@book{chen1984linear,
   author = {Chi-Tsong. Chen},
   isbn = {978-0-03-071691-1},
   publisher = {Saunders College Publishing},
   title = {Linear system theory and design},
   year = {1984}
}

@inproceedings{bonnabel2009invariant,
   author = {Silvere Bonnabel and Philippe Martin and Erwan Salaun},
   doi = {10.1109/CDC.2009.5400372},
   isbn = {978-1-4244-3871-6},
   booktitle = {IEEE Conference on Decision and Control (CDC)},
   month = {12},
   pages = {1297-1304},
   publisher = {IEEE},
   title = {Invariant Extended Kalman Filter: theory and application to a velocity-aided attitude estimation problem},
   year = {2009}
}

@article{hamel2017position,
   author = {Tarek Hamel and Claude Samson},
   doi = {10.1016/j.automatica.2017.04.045},
   issn = {00051098},
   journal = {Automatica},
   month = {8},
   pages = {137-144},
   title = {Position estimation from direction or range measurements},
   volume = {82},
   year = {2017}
}

@inproceedings{thienel2001coupled,
  title={A coupled nonlinear spacecraft attitude controller/observer with an unknown constant gyro bias},
  author={Thienel, J and Sanner, Robert M},
  booktitle={Proceedings of the 40th IEEE Conference on Decision and Control},
  volume={4},
  pages={3441--3446},
  year={2001},
  organization={IEEE}
}

@article{grip2011attitude,
  title={Attitude estimation using biased gyro and vector measurements with time-varying reference vectors},
  author={Grip, H{\aa}vard Fj{\ae}r and Fossen, Thor I and Johansen, Tor A and Saberi, Ali},
  journal={IEEE Transactions on automatic control},
  volume={57},
  number={5},
  pages={1332--1338},
  year={2011},
  publisher={IEEE}
}

@article{laidig2021broad,
  title={BROAD—A benchmark for robust inertial orientation estimation},
  author={Laidig, Daniel and Caruso, Marco and Cereatti, Andrea and Seel, Thomas},
  journal={Data},
  volume={6},
  number={7},
  pages={72},
  year={2021},
  publisher={MDPI}
}

@article{Melis2026,
    author ={ Melis, Alessandro and Bouazza, Tarek and Alnahhal, Hassan and Benahmed, Siffedine and Berkane, Soulaimane and Hamel, Tarek},
    title = {Scalar-measurement attitude estimation on SO(3) with bias compensation},
    journal = {ICRA},
    year = {2026}
}

\end{document}